\documentclass[aps,longbibliography,showpacs,twocolumn,superscriptaddress]{revtex4-1}
\usepackage{amsmath,amssymb,amsfonts,bm}
\usepackage{graphicx}
\usepackage{epstopdf}
\usepackage{dcolumn}
\usepackage{mathrsfs}
\usepackage{bbold}
\usepackage{dsfont}
\usepackage{float}
\usepackage[colorlinks=true,linkcolor=magenta,citecolor=magenta, urlcolor=magenta,bookmarks=false]{hyperref}
\usepackage{tgtermes}
\usepackage{multirow}
\usepackage{bbm}
\usepackage{wasysym}
\usepackage{newtxtext}
\usepackage[varvw]{newtxmath}
\usepackage{hyperref}
\usepackage{array}
\usepackage{bm}
\usepackage{color}
\usepackage{float}
\usepackage{graphicx}
\usepackage{natbib}
\usepackage{newtxtext}
\usepackage{newtxmath}
\usepackage[caption=false]{subfig}
\usepackage{mathrsfs}
\usepackage{amsmath,amssymb}
\usepackage{verbatim}
\usepackage{amsfonts}
\usepackage{dcolumn}
\usepackage{ulem}

\newcommand{\beq}{\begin{eqnarray} }
\newcommand{\eeq}{\end{eqnarray} }
\newcommand{\Beq}{\begin{eqnarray*} }
\newcommand{\Eeq}{\end{eqnarray*} }
\newcommand{\Bmat}{\left(\begin{matrix}}
\newcommand{\Emat}{\end{matrix}\right)}
\newcommand{\up}{\uparrow}
\newcommand{\dn}{\downarrow}

\begin{document}

\title{Multinode quantum spin liquids in extended Kitaev honeycomb models: the view from variational Monte Carlo}

\author{Jiucai Wang}
\email{jcwphys@hku.hk}
\affiliation{Department of Physics and HKU-UCAS Joint Institute for Theoretical and Computational Physics, The University of Hong Kong, Pokfulam Road, Hong Kong, China}
\affiliation{HK Institute of Quantum Science \& Technology, The University of Hong Kong, Pokfulam Road, Hong Kong, China}

\author{B. Normand}
\email{bruce.normand@psi.ch}
\affiliation{PSI Center for Scientific Computing, Theory and Data, CH-5232 Villigen-PSI, Switzerland}
\affiliation{Institute of Physics, Ecole Polytechnique F\'ed\'erale de Lausanne (EPFL), CH-1015 Lausanne, Switzerland}

\author{Zheng-Xin Liu}
\email{liuzxphys@ruc.edu.cn}
\affiliation{School of Physics and Beijing Key Laboratory of Opto-electronic Functional Materials and Micro-nano Devices, \\ 
Renmin University of China, Beijing, 100872, China}
\affiliation{Key Laboratory of Quantum State Construction and Manipulation (Ministry of Education), \\ 
Renmin University of China, Beijing, 100872, China}

\date{\today}

\begin{abstract}
We discuss the discovery by variational Monte Carlo (VMC) methods of a series of multinode quantum spin liquids (QSLs) in extended Kitaev models on the honeycomb lattice. Like the gapless Kitaev spin liquid with its two nodes at K and K$^\prime$, these multinode QSLs are characterized by an emergent Z$_2$ gauge structure and a discrete number of symmetry-protected Majorana cones in their low-energy excitation spectrum. Because the cones are gapped by weak magnetic fields, nonzero Chern numbers are obtained and the ground state becomes one of many possible Abelian or non-Abelian chiral spin liquids. Here we focus on the projective symmetry group (PSG)-guided VMC approach to the Kitaev model with various symmetry-allowed extended interactions. Based on the VMC phase diagrams of these models, we propose a framework for the classification of nodal QSLs that includes the PSG, the chiralities of the cones, and the way in which the cones are symmetry-related. At present, the known candidate Kitaev materials seem to lie outside the parameter regimes of the multinode QSL phases. However, with more than 100 Z$_2$ PSGs for spin-orbit-coupled states on the honeycomb lattice, we anticipate that more than one multinode QSL will be realized experimentally in future work.
\end{abstract}

\maketitle

\section{Introduction}

Quantum spin liquids (QSLs) are exotic phases of magnetic matter that avoid any form of symmetry-breaking long-range order even at zero temperature \cite{Balents2016,zhouyi,Yu2019,Senthil2020}. QSLs result from strong quantum fluctuations in the site-spin basis, and are characterized not by any local order parameter but instead by their long-range quantum entanglement, intrinsic fractionalized spin excitations, and emergent gauge fields \cite{Levin2006,Kitaev2006}. However, because most quantum spin systems in two or higher dimensions order magnetically at low temperatures, the realization of QSL ground states remains a challenge both in lattice models and in quantum magnetic materials. 

The most significant progress in the field of QSLs to date in the 21st century has been the proposal of the exactly soluble spin-1/2 Kitaev honeycomb model \cite{Kitaev}, of mutually frustrating bond-dependent nearest-neighbor Ising interactions, which contains QSL ground states with both gapped and gapless excitations. The gapped Kitaev QSL has Z$_2$ Abelian topological order, while the quintessential ``Kitaev spin liquid'' (KSL) is the gapless state, whose low-energy excitations form two Majorana cones, whereas its Z$_2$ flux excitations are gapped. In an applied magnetic field, these Majorana cones become gapped and the resulting state is a chiral spin liquid (CSL) with Ising-type non-Abelian anyonic excitations, which have possible applications in topologically protected quantum computation \cite{Nayak2008}. The KSL is the paradigm for a QSL state quite different from the resonating-valence-bond (RVB) QSLs proposed for systems dominated by Heisenberg interactions \cite{Anderson1973}, due among multiple factors to the lowering of the spin symmetry from continuous to discrete.

A natural issue is to realize the KSL in quantum magnetic materials. Kitaev-type interactions were shown \cite{rjk2009,rcjk2010} to emerge in low-spin 4d or 5d transition-metal ions, which combine substantial spin-orbit coupling (SOC), modest electronic correlations, and an edge-sharing ligand geometry. This moved the experimental realization of the Kitaev honeycomb model to the forefront of research in strongly correlated materials \cite{Knolle2018,Takagi2019,Motome2020,Trebst2022}. Although bond-dependent Kitaev interactions can be observed directly in polarized resonant X-ray data \cite{Kim2015}, the ``candidate Kitaev materials'' Na$_2$IrO$_3$ \cite{singh,XLiu,ryea,rchoietal} and $\alpha$-RuCl$_3$ \cite{rsea,rjea,Banerjee2016,rcaoetal,Banerjee2017} also possess significant non-Kitaev interactions that lead to low-temperature magnetic order.

However, this order is rather fragile in an applied magnetic field \cite{Kindo2015,Yu2017,Sun2018}, giving way in $\alpha$-RuCl$_3$ to a disordered phase with the characteristics of a QSL. These include a broad continuum of spin excitations observed at high energies by inelastic neutron scattering (INS) \cite{Nagler2018}, low-energy excitations observed by nuclear magnetic resonance (NMR) that are gapless for an in-plane field \cite{YuW2017} and gapped with an out-of-plane field component \cite{Baek2017}, an excitation continuum in low-temperature Raman scattering \cite{Burch2015}, and reports of a half-integer-quantized plateau in the thermal Hall signal over a small range of temperature and field \cite{Matsuda2018,Matsuda2021}. However, as with candidate materials for realizing QSL phases by other mechanisms, it is important to exclude the possibility that the magnetically disordered state is a consequence of extrinsic factors, such as stacking faults \cite{rcaoetal,Dinnebier2017} or quenched disorder \cite{Kitagawa2018,Loidl2020,Choi2023}. Thus the possibility of a field-induced disordered phase remains the target of intensive current research \cite{Tanaka2020,Ong2021,Ong2022,Bruin2022,Zhou2023}.

From a theoretical standpoint, non-Kitaev interactions including the Heisenberg ($J$), off-diagonal symmetric $\Gamma$ and $\Gamma'$ interactions appear at the same order as the Kitaev interaction ($K$) in a strong-coupling treatment \cite{Kee2024}. This class of interactions establishes a well-defined parameter space in which to interpret ``proximate Kitaev'' physics in candidate materials, including the observed zigzag ground state and possible field-induced QSL phases. The very large phase space offers many possibilities for unconventional phases of quantum magnetic matter lying beyond the known materials, and systematic theoretical studies are required to explore it.

\begin{figure}[t]
\centering
\includegraphics[width=\linewidth]{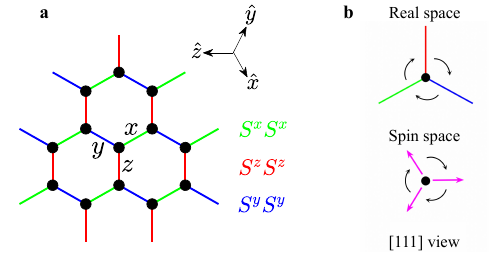}
\caption{{\bf a} Schematic representation of the honeycomb lattice with nearest-neighbor bond types $\gamma = x$, $y$, and $z$. The Kitaev model involves only Ising interactions of the $\gamma$ component of the $S = 1/2$ spins on bond $\gamma$. 
{\bf b} The strong spin-orbit coupling in candidate Kitaev materials ensures a discrete symmetry group, for example the three-fold rotational symmetry around the $[111]$ axis (the $\hat n_c$ direction) both in real space and in spin space.}
\label{lattice}
\end{figure}

In this article we discuss one of the most exciting possibilities to have emerged to date, the multinode QSL states. These can be considered as a generalization of KSL physics in which the mechanism for energy minimization in the presence of the competing extended interactions remains a bandstructure of quasi-free spinons with point nodes in two dimensions, but with no strong distinction between Majorana and flux degrees of freedom. We first show how the investigation of extended Kitaev models on the honeycomb lattice is made both more systematic and more straightforward by applying the projective symmetry group (PSG). In a PSG-guided variational Monte Carlo (VMC) approach, we compute phase diagrams in the presence of different extended interactions to find that the ``generic'' KSL is accompanied by several multinode Z$_2$ QSLs, which can be characterized by the number of symmetry-protected cones in their low-energy excitation spectrum. By extending this study to finite fields, we demonstrate that different CSLs are obtained that realize different members of Kitaev's 16-fold way classification. We propose that a complete classification of the nodal QSLs allowed with a given combination of spin and spatial symmetry should include not only the PSG but also information about the cones, specifically their total number, chiralities, and how they are related by the symmetries. This observation can be applied to classify nodal QSLs on different lattices and with different extended symmetry groups.

This article is organized as follows. We begin in Sec.~\ref{SecII} by defining the extended honeycomb Kitaev models we consider. In Sec.~\ref{SecIII} we summarize the VMC method, which is based on PSG-guided, Gutzwiller-projected wavefunctions. We discuss in Sec.~\ref{SecIV} the quantum phase diagrams of our extended Kitaev models, illustrate the physical properties of the three previously unknown multinode QSLs they contain, and show how these QSLs are related. In Sec.~\ref{SecV} we turn to the response of multinode QSLs to applied magnetic fields, with particular focus on the field-induced CSL states that possess new topological properties. In Sec.~\ref{SecVI} we discuss the properties of Majorana cones and the classification of multinode QSLs, before concluding in Sec.~\ref{SecVII} by summarizing these results and their more general context.

\section{Extended Kitaev models}
\label{SecII} 

The Kitaev model is a compass model on the honeycomb lattice \cite{Kitaev,Brink2015}, which to establish notation we express as $H_K = \sum_{\langle i,j \rangle \in\gamma} \, K \, S_i^\gamma S_j^\gamma$. Here $\langle i,j \rangle$ denotes nearest-neighbor sites, $\gamma$ labels the bond type $\langle i,j \rangle$ on the honeycomb lattice, and $\alpha, \beta, \gamma$ denote the spin index, with $S_i^\gamma S_j^\gamma$ being an Ising interaction on bond $\gamma$ (Fig.~\ref{lattice}a). The model is made more symmetric by identifying the normal to the honeycomb plane as $\hat n_c = {1\over\sqrt3}(\hat x + \hat y + \hat z)$ and the orientation of the $\gamma$-bond as $\hat n_\gamma = {1\over\sqrt2}(\hat\alpha - \hat\beta)$, where $\hat x, \hat y, \hat z$ are the axes in the spin frame. In this basis, the Kitaev model has point-group symmetry $D_{3d} \times Z_2^T$, with $Z_2^T = \{E, T\}$ the time-reversal group \cite{YouPSG}. A $C_3$ rotation about $\hat n_c$ then permutes the $x,y,z$ bonds and the corresponding spin directions simultaneously (Fig.~\ref{lattice}b). 

\begin{figure}[t]
\includegraphics[width=\linewidth]{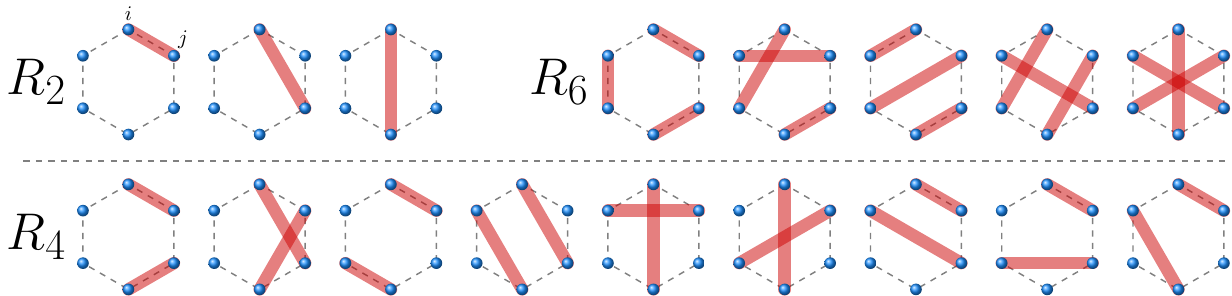}
\caption{Schematic representation of the 2-spin ($R_2$, left upper row), 6-spin ($R_6$, right upper row), and 4-spin ($R_4$, bottom row) interactions on each plaquette of the honeycomb lattice \cite{JCWang2023}. In $R_2$, the numbers of each element are respectively 6, 6, and 3. In $R_4$, they are respectively 6, 6, 3, 3, 6, 3, 6, 6, and 6. In $R_6$, they are respectively 2, 6, 3, 3, and 1. Red shaded bars connecting one pair of sites ($i,j$) represent the Heisenberg interaction $\vec S_i \! \cdot \! \vec S_j$. The $R_4$ and $R_6$ interactions have a sign structure of $+1$ ($-1$) for an even (odd) number of transpositions of the bars.}
\label{Ring}
\end{figure}

The most general model preserving these symmetries contains four different types of interaction on the nearest-neighbor bonds, 
\begin{eqnarray}\label{GenericKitaev}
H & = & \!\!\! \sum_{\langle i,j \rangle \in \alpha \beta (\gamma)} \!\!\! K \, S_i^\gamma S_j^\gamma + J \, {\vec S}_i \! \cdot \! {\vec S}_j + \Gamma \, \big(S_i^\alpha S_j^\beta + S_i^\beta S_j^\alpha \big) \nonumber \\
& & \quad + \,\, \Gamma' \, \big(S_i^\alpha S_j^\gamma + S_i^\gamma S_j^\alpha + S_i^\beta S_j^\gamma + S_i^\gamma S_j^\beta \big) \, + \, \dots, 
\end{eqnarray}
where the ellipsis denotes symmetry-allowed interactions on further-neighbor bonds. For candidate Kitaev materials, this generic model can also be derived microscopically for spin-orbit-entangled $d$-electrons with $J_{\rm eff} = 1/2$ moments \cite{c1,Khaliullin2022}.

Here we will also consider the leading correction to the nearest-neighbor bilinear Hamiltonian [Eq.~\eqref{GenericKitaev}] in the strong-coupling expansion of the Hubbard model \cite{Yang2010,Yang2012}, which is the ring-exchange interaction on the hexagon \cite{Motrunich2005}
\begin{eqnarray}
H_{\rm Ring} \, = \, \sum_{\hexagon} \, J_{\rm R} \, \hat P_{\hexagon}, 
\end{eqnarray}
where $\hat P_{\hexagon} = - {3\over 8} + R_2 + R_4 + R_6$ with $R_2$, $R_4$, and $R_6$ representing respectively the 2-spin, 4-spin, and 6-spin interactions shown in Fig.~\ref{Ring}. Despite the high-order nature of this interaction, phases induced by ring-exchange have been invoked for the candidate Kitaev material Na$_2$Co$_2$TeO$_6$ \cite{Janssen2023}. 

\section{PSG-guided VMC method}
\label{SecIII} 

The Kitaev model is exactly soluble in the Majorana representation of the spin, $S^\gamma_i = \frac{i}{2} b_i^\gamma c_i$ ($\gamma = x, y, z$), with the constraint $b_i^x b_i^y b_i^z c_i = 1$. The $c$ fermions propagate freely while the $b^\gamma$ fermions are localized on bond $\gamma$ and couple to the $c$-fermions by providing a background Z$_2$ gauge field. For isotropic coupling constants, the spectrum is gapless with two cones located at K and K$^\prime$ points of the Brillouin zone (BZ). The KSL ground state is obtained after projecting to the physical Hilbert space.

In contrast to the exact solubility of the pure Kitaev model, the generic extended Kitaev model [Eq.~\eqref{GenericKitaev}] can be treated only approximately. In the absence of a small parameter to control an analytical perturbative description, much of the progress made in understanding such strongly frustrated and correlated systems relies on numerical methods. However, exact diagonalization (ED) is limited by the system size, quantum Monte Carlo methods suffer from the ``sign problem'' \cite{Sandvik2010}, and renormalization-group methods such as density matrix renormalization group (DMRG) \cite{DMRG2005} are limited to very narrow cylinders, while tensor RG methods \cite{Cirac2021} are limited for gapless systems and by their in bond dimension. Thus we pursue the variational Monte Carlo (VMC) method, which has also brought valuable insight into the study of strongly correlated many-body systems \cite{Gros2007}, and specifically we introduce a PSG-guided VMC approach that (i) identifies the appropriate low-energy effective theory and (ii) describes nodal QSLs quite readily.

For an overview of the VMC method, we first introduce the fermionic slave-particle representation $S_i^m = \frac{1}{2} C_i^\dagger \sigma^m C_i$, where $C_i = (c_{i\uparrow}, c_{i\downarrow})^{\rm T}$, $m \equiv x,y,z$, and $\sigma^m$ are Pauli matrices. This complex fermion representation is equivalent to the Majorana representation introduced by Kitaev. The local particle-number constraint takes the form $\hat{N_i} = c_{i\uparrow}^\dagger c_{i\uparrow} + c_{i\downarrow}^\dagger c_{i\downarrow} = 1$, and imposing it rigorously at every site ensures that the Hilbert space of the fermions has the same size as that of the physical spin. Introducing the partner of $C_i$ under time-reversal and particle-hole transformation, $\bar C_i = (c_{i\dn}^\dag , -c_{i\up}^\dag)^{\rm T}$, allows the spin operator to be conveniently expressed as $S^m_i = {\textstyle \frac{1}{4}} {\rm Tr}(\psi_i^{\dagger} \sigma^m \psi_i)$, with $\psi_i = (C_i, \bar C_i)$. This quantity has both an SU(2) spin-rotation symmetry and an independent local SU(2) symmetry that can be considered as a gauge structure.

Quite generally, the spin interactions can be reexpressed in terms of these fermionic operators and then decoupled into a noninteracting mean-field Hamiltonian, $H_{\rm mf}^{\rm SL}$. While any QSL ground state preserves the full symmetry group $G$, a general symmetry operation of $H_{\rm mf}^{\rm SL}$ is a space-time and spin operation in $G$ combined with an SU(2) gauge transformation. The larger group formed by these symmetry operations is the PSG \cite{igg}. In the Methods section we show explicitly how the PSG is applied to the KSL \cite{YouPSG} and to the most general spin-orbit-coupled QSL Ansatz for the $K$-$J$-$\Gamma$ model with only nearest-neighbor interactions \cite{JCWang2019}. The major reduction in the number of independent model parameters achieved through its application is the reason why the PSG has become an important tool in studying QSLs. 

In addition to $H_{\rm mf}^{\rm SL}$, in our trial states we allow competing magnetically ordered phases by including the decouplings $H_{\rm mf}^{\rm Order} = \frac{1}{2} \sum_i \pmb M_i \! \cdot \! C_i^\dag\pmb \sigma C_i$, with $\pmb M_i$ the pattern of ordered moments. We then use VMC to perform Gutzwiller projection on the mean-field ground states, $|\Psi_{\rm mf} (\pmb R)\rangle$, to enforce the particle-number constraint. The projected states $|\Psi (\pmb R) \rangle = P_{\rm G} |\Psi_{\rm mf}(\pmb R) \rangle$ provide variational wavefunctions, where $\pmb R$ denotes the variational parameters. The energy of the trial state, $E (\pmb R) = \langle \Psi (\pmb R) |H| \Psi (\pmb R) \rangle / \langle \Psi (\pmb R)| \Psi (\pmb R) \rangle$, is computed by Monte Carlo sampling and the optimal set $\pmb R$ is determined by minimizing $E(\pmb R)$; further details are provided in the Methods section. 

\section{Multinode QSLs}
\label{SecIV} 

In most known candidate Kitaev materials, the Kitaev interaction is found to be ferromagnetic ($K < 0$), while the Heisenberg ($J$) interactions are thought to extend to third neighbors and take both signs \cite{Winter2016,Winter2017}. There is less debate about the $\Gamma$ term, which must be large and positive ($\Gamma > 0$) to explain the sizeable dependence on field orientation of the anisotropic magnetic susceptibility in $\alpha$-RuCl$_3$ \cite{Janssen2017,Sears2020}. To control the parameter space of our studies, we focus on interactions satisfying $K < 0$, $\Gamma > 0$, and $J_{\rm R} > 0$, while we allow $J$ and $\Gamma'$ to be both positive and negative. In this section we also set the applied magnetic field to zero and present detailed VMC results for several extended Kitaev models \cite{JCWang2023,JCWang2019,JCWang2020}, with particular focus on the multinode QSL phases they contain. 

\begin{figure*}[t]
\centering
\includegraphics[width=\linewidth]{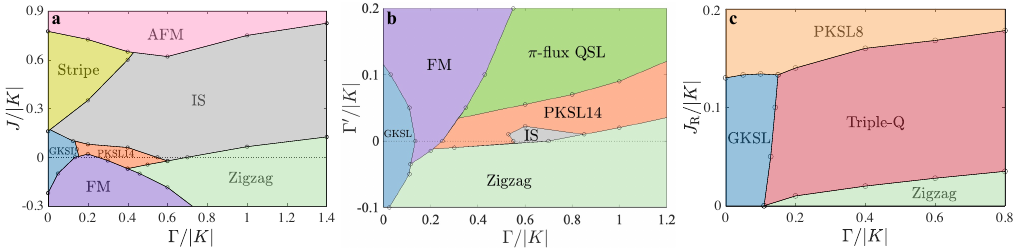}
\caption{Phase diagrams of three extended Kitaev honeycomb models obtained from VMC calculations extrapolated to large system size.
{\bf a} Phase diagram of the quantum $K$-$J$-$\Gamma$ model for $K < 0$, $\Gamma >0$ \cite{JCWang2019}.
{\bf b} Phase diagram of the quantum $K$-$\Gamma$-$\Gamma'$ model for $K < 0$, $\Gamma >0$ \cite{JCWang2020}.
{\bf c} Phase diagram of the quantum $K$-$\Gamma$-$\Gamma'$-$J_{\rm R}$ model for $K < 0$, $\Gamma >0$, $J_{\rm R}>0$, and $\Gamma'/|K| = -0.05$ \cite{JCWang2023}. Abbreviations: GKSL, generic Kitaev spin liquid; PKSL14, proximate Kitaev QSL with 14 cones; PKSL8, proximate Kitaev QSL with 8 cones; $\pi$-flux QSL, QSL with a distinct PSG (including translation-symmetry fractionalization); IS, incommensurate spiral order; AFM, antiferromagnetic order; FM, ferromagnetic order.}
\label{Fig_phsdgrm}
\end{figure*}

\begin{figure*}[t]
\centering
\includegraphics[width=\linewidth]{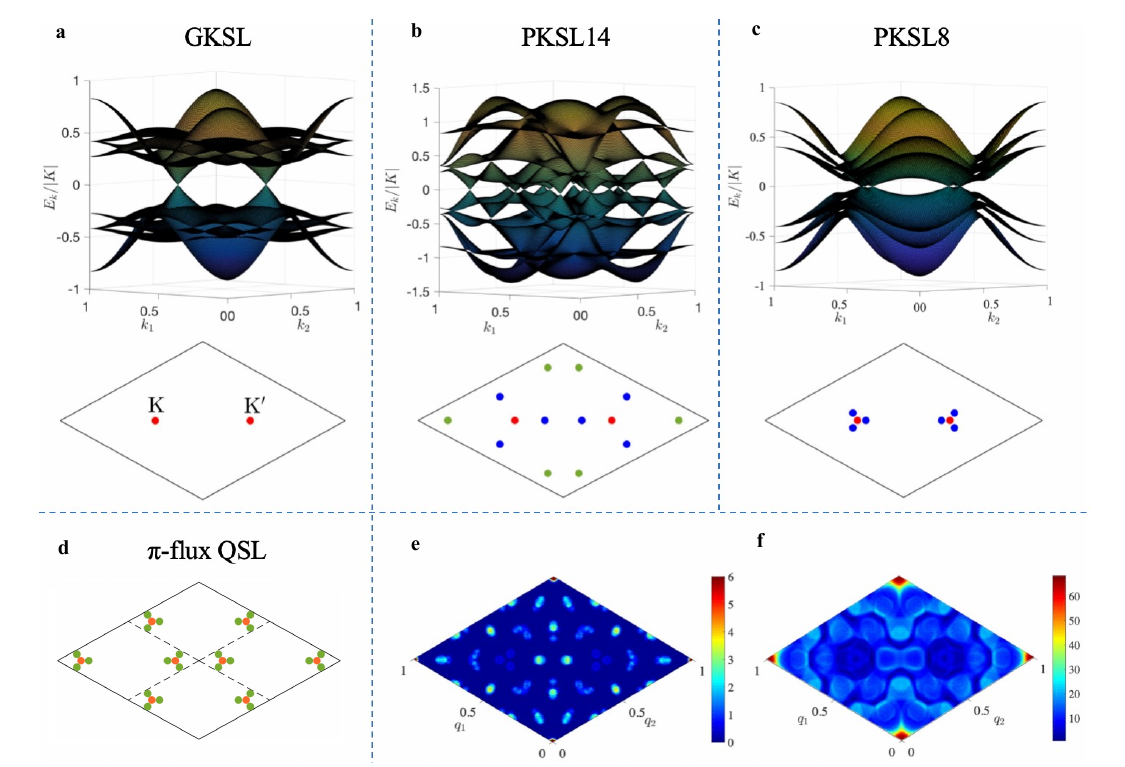}
\caption{{\bf a} Spinon dispersion in the GKSL, drawn with $\Gamma/|K| = 0.1$ and $J = 0$, showing 2 cones. 
{\bf b} Spinon dispersion in the PKSL14, drawn with $\Gamma/|K| = 0.3$ and $J = 0$, showing 14 cones \cite{JCWang2019}. 
{\bf c} Spinon dispersion in the PKSL8, drawn with $\Gamma/|K| = 0.1$, $\Gamma'/|K| = - 0.05$, and $J_{\rm R}/|K| = 0.15$, showing 8 cones \cite{JCWang2023}. The positions of the nodes of these cones in the BZ are depicted separately below each panel; nodes of the same color are symmetry-related.  
{\bf d} Positions of the 32 nodes in the $\pi$-flux QSL state \cite{JCWang2020}. 
{\bf e} DSF of the PKSL14 at low energies (integrated over the range $0 \le \omega/|K| \le 0.08$), computed at the mean-field level. 
{\bf f} Mean-field DSF of the PKSL14 at intermediate energies (integrated over $0.13 \le \omega/|K| \le 0.17$) \cite{JCWang2019}.}
\label{Fig_dispersions}
\end{figure*}

\subsection{GKSL and PKSL14 in the $\pmb K$-$\pmb J$-$\pmb\Gamma$ model} 

The phase diagram of the quantum $K$-$J$-$\Gamma$ model computed by VMC is shown in Fig.~\ref{Fig_phsdgrm}a \cite{JCWang2019}. Although a number of QSL phases appear at the mean-field level, only two remain after the VMC projection to the physical Hilbert space. One is the generic KSL (GKSL), which extends beyond the Kitaev point to approximately $|J/K| = 0.2$ at $\Gamma = 0$ and $\Gamma/|K|$ = 0.15 at $J = 0$. The phase boundary along the Heisenberg-Kitaev axis agrees with the result of Ref.~\cite{HCJiang}, and both boundaries quantify the considerations of Ref.~\cite{rsyb}. Beyond the GKSL on the $\Gamma/|K|$ axis we find the only other proximate QSL in this model, which we name the PKSL14. Its most striking property is that its spinon dispersion has 14 cones in the first BZ, which we illustrate in Fig.~\ref{Fig_dispersions}b by comparison with the familiar pair of cones exhibited by the GKSL (Fig.~\ref{Fig_dispersions}a). The PKSL14 therefore provides a clear statement that multinode QSLs, meaning states that optimize the energy of fractional spin entities by a generalization of the nodal physics of the KSL, are highly competitive candidate ground states in proximate Kitaev models. We stress that, despite their clear differences, the GKSL and PKSL14 have the same PSG, which underlines the fact that a full classification of multinode QSLs must also require information about the cones. As we discuss below, all of the cones in any such QSL are robust against local perturbations due to the combination of time-reversal $(T)$ and spatial inversion $(P)$ symmetries, which also give them explicit inter-cone symmetry relations. 

For perspective we note that most of the phase diagram of Fig.~\ref{Fig_phsdgrm}a is occupied by the five magnetically ordered states known from the classical $K$-$J$-$\Gamma$ model \cite{c1}, among which the only surprise is the persistence of incommensurate order in a spin-1/2 model over such a wide parameter regime. Only at small $J$ do the classical and quantum models differ significantly, but we remark in this context that zigzag magnetic order is favored even at $J = 0$ for all parameter ratios $\Gamma/|K| \gtrsim 0.6$, including in the pure $S = 1/2$ $\Gamma$ model (where a zigzag phase is confirmed by matrix product state (MPS) and tensor-network methods \cite{pqgm}). Because these results contradict an infinite DMRG (iDMRG) study \cite{c3} that reported only QSL phases over the entire $\Gamma/|K|$ axis, it was shown explicitly by VMC on very narrow systems equivalent to the iDMRG cylinder that indeed a QSL phase persists, which should be regarded as an expression of 1D physics \cite{JCWang2019}. As intermediate steps to the quantitative phase diagram of the infinite 2D quantum $K$-$\Gamma$ model, we remark that DMRG studies of the $K$-$\Gamma$-$\Gamma'$ model (Sec.~\ref{SecIV}B) have also found a proximate QSL distinct from the KSL, which was named the K$\Gamma$SL \cite{Gordon2019}, as has a 2D mean-field Ansatz \cite{Yilmaz2022}, but a final phase diagram remains the target of ongoing research \cite{Kee2024,Buessen2021,QLuo2021}. 

Turning to the nature of the phase transitions, it was found that every single line in Fig.~\ref{Fig_phsdgrm}a is ﬁrst-order, with abrupt level crossings in the ground-state energies and discontinuities in the optimal variational parameters. This statement applies not only to transitions between ordered states but also at the QSL-order boundaries and at the QSL-QSL transition, as shown explicitly in Ref.~\cite{JCWang2019}. Thus it appears empirically that the possibility of continuity, or of coexistence between magnetic and Z$_2$ topological order, is hard to realize in models dominated by discrete symmetries. 

To describe experiments probing the spin response of proximate Kitaev materials, we recall that the gap of the $b^\gamma$ fermions in the KSL ensures that the spin excitations are also gapped, despite the gapless nature of the $c$ fermions. This physics persists at the qualitative level in the GKSL (Fig.~\ref{Fig_dispersions}a), in that the $c$-$b^\gamma$ hybridization remains very weak at low energies and thus the gapless spin excitations that are permitted when both $J$ and $\Gamma$ are finite have only minimal spectral weight around K and K$^\prime$. The PKSL14 is quite different: when expressed in the Majorana basis, its spinons are strongly hybridized $c$-$b^\gamma$ combinations at all energies and wavevectors, leading to a gapless spin response with strong low-energy spectral weight. In Fig.~\ref{Fig_dispersions}e we show that the dynamical structure factor (DSF), $S({\pmb q},\omega)$, contains low-energy spin excitations whose origin lies in both intra- and intercone spinon processes, as one may read from the locations of the nodes in the BZ (Fig.~\ref{Fig_dispersions}b). Once the energy transfer exceeds the height of the cones in the spinon dispersion of Fig.~\ref{Fig_dispersions}b, $S({\pmb q},\omega)$ exhibits a very rich structure (Fig.~\ref{Fig_dispersions}f). 

We remark here that the DSFs shown in Figs.~\ref{Fig_dispersions}e-f were computed for the unprojected fermionic states, as only at the mean-ﬁeld level can the calculations be performed for large systems to obtain high $\pmb q$-resolution. The calculation of $S(\pmb q,\omega)$ with the Gutzwiller projection is possible only for very small systems that cannot capture the detailed node structure of a multinode QSL. However, because the Z$_2$ gauge ﬁeld is deconﬁned in the QSL phase, interactions between spinons are rather weak, and hence the structure of the cones is expected to remain after Gutzwiller projection, making the results we show a good qualitative description of the full DSF. From the spinon dispersion relations in any multinode QSL, and in particular the GKSL, we make an important remark: finding a low-energy signal by INS, and also by thermodynamic measurements, may be very difficult due to the low density of states in the cone regions, possibly resulting in the erroneous identification of a gapped QSL, whereas an accurate identification of a gapless nodal state may require NMR or electron spin resonance (ESR) measurements. 

\subsection{PKSL14 and $\pmb\pi$-flux QSL in the $\pmb K$-$\pmb\Gamma$-$\pmb\Gamma'$ model} 

In Fig.~\ref{Fig_phsdgrm}b we set $J = 0$ to minimize ordering tendencies and show the VMC phase diagram of the quantum $K$-$\Gamma$-$\Gamma'$ model \cite{JCWang2020}. Again the GKSL is destabilized by off-diagonal terms ($\Gamma$ and $\Gamma'$) approximately 10\% of $|K|$. Our results confirm that small negative $\Gamma'$ interactions stabilize the zigzag ordered phase \cite{Gordon2019,tensor}, which is valuable for comparison with experimental observations. From the standpoint of multinode QSLs, however, positive $\Gamma'$ interactions have a dramatic effect in stabilizing the PKSL14, which at $\Gamma'/|K| = 0.05$ extends over a very wide range of $\Gamma/|K|$. Here one may consider that this enhanced stability translates to a larger energy range of cone-type low-energy spinon dispersions (Fig.~\ref{Fig_dispersions}b), improving the visibility of the PKSL14 in spectroscopic experiments, as well as of its field-induced phase transitions (Sec.~\ref{SecV}). 

At still larger positive $\Gamma'$ we find an even more exotic QSL, with 32 cones in the first BZ \cite{JCWang2020}. This we name a symmetric $\pi$-flux QSL state after its PSG, which is different from that of the GKSL and PKSL14. The 32 nodes can be characterized as 8 cones in the compact BZ, as shown in Fig.~\ref{Fig_dispersions}d, an enhanced periodic structure visible in the spinon dispersions that indicates a fractionalization of the translational symmetry. Like the PKSL14, the $\pi$-flux state has a gapless spin response. For completeness we note again that all transitions in the phase diagram of the $K$-$\Gamma$-$\Gamma'$ model (Fig.~\ref{Fig_phsdgrm}b) are of first order, including the boundary between the PKSL14 and the $\pi$-flux state. 

\begin{figure}[t]
\centering
\includegraphics[width=\linewidth]{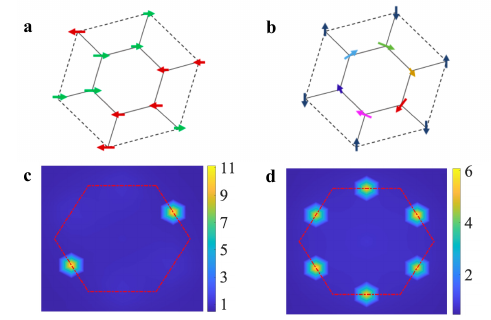}
\caption{{\bf a} Schematic representation of zigzag magnetic order.
{\bf b} Schematic representation of triple-$\bf Q$ order, which is formed by the vector sum of three C$_3$-related zigzag configurations.
{\bf c} Static spin structure factor, $S(\pmb{q})$, for the zigzag ordered phase (calculated with $\Gamma/|K| = 0.4$, $\Gamma'/|K| = - 0.05$).
{\bf d} $S(\pmb{q})$ for the triple-$\bf Q$ ordered phase (calculated with $\Gamma/|K| = 0.4$, $\Gamma'/|K| = - 0.05$, and $J_{\rm R}/|K| = 0.1$) \cite{JCWang2023}. The red dotted line represents the first crystallographic BZ and the bright points in the figures are the M points.}
\label{ZZPattern}
\end{figure}

\subsection{PKSL8 in the $\pmb K$-$\pmb\Gamma$-$\pmb\Gamma'$-${\pmb J}_{\rm R}$ model} 

Introducing the ring-exchange interaction, $J_{\rm R}$, makes it difficult to explore the full phase space of the extended model. In Fig.~\ref{Fig_phsdgrm}c we show the VMC phase diagram of the quantum $K$-$\Gamma$-$\Gamma'$-$J_{\rm R}$ model with the $\Gamma'$ interaction set to $-0.05|K|$ \cite{JCWang2023}. Although this value was chosen to reinforce the experimentally observed zigzag-ordered ground state (Fig.~\ref{Fig_phsdgrm}b) \cite{JCWang2020,Gordon2019,tensor}, rather than to maximize the chances of finding multinode QSLs, the results nonetheless contain two important pieces of insight. 

Here again the GKSL is stable up to $J_{\rm R}/|K| = 0.13$. However, the QSL proximate to the GKSL is a PKSL8, an eight-cone state whose spinon dispersions and node locations are shown in Fig.~\ref{Fig_dispersions}c. Its four pairs of cones include one pair at K and K$^\prime$ points, and three cones very close to either K or K$^\prime$ that are connected by $C_3$ rotation, and whose pair structure is revealed under inversion or time-reversal symmetry. The PKSL8 shares the same PSG as the KSL and PKSL14, although again the reciprocal-space structure of its spinon bands and gapless spin excitations is quite different from both. Again we find from VMC that the GKSL-PKSL8 phase transition is sharply first-order. We remark that adding a $J_{\rm R}$ term to a $K$-$\Gamma$-$\Gamma'$ model with $\Gamma'/|K| = 0.05$, i.e.~to a system of PKSL14 and $\pi$-flux ground states (Fig.~\ref{Fig_phsdgrm}b), might have significant potential to reveal new multinode QSLs. 

In addition to the QSLs, Fig.~\ref{Fig_phsdgrm}c contains two magnetically ordered phases, the zigzag state (Fig.~\ref{ZZPattern}a) and a triple-$\bf Q$ state (Fig.~\ref{ZZPattern}b) \cite{JCWang2023}. This latter has an 8-site magnetic unit cell and a $C_3$ rotation symmetry that includes both spin and lattice rotation (Fig.~\ref{lattice}b), and corresponds to a superposition of three symmetry-related zigzag order parameters \cite{LiYuan2021,LiYuan2022}, as shown by the static structure factors displayed in Figs.~\ref{ZZPattern}c and \ref{ZZPattern}d. A positive $J_{\rm R}$ therefore appears to drive quantum fluctuations that restore $C_3$ symmetry, and the resulting triple-$\bf Q$ state is adjacent to two QSLs (Fig.~\ref{Fig_phsdgrm}c). This suggests a role as a phase intermediate between the zigzag phase and the QSL regime, and hence as an indicator in the experimental search for QSLs. 

\begin{figure*}[t]
\centering
\includegraphics[width=\linewidth]{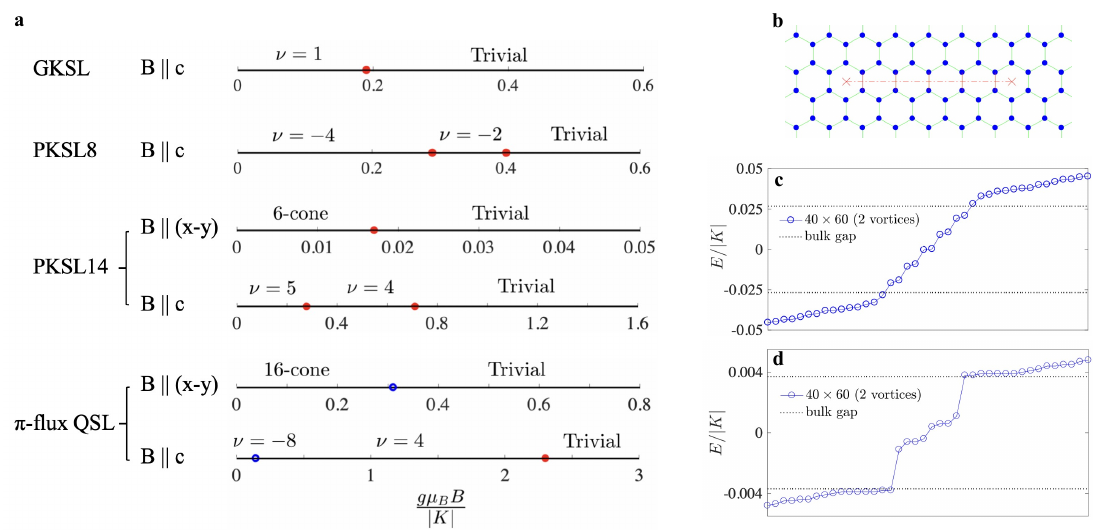}
\caption{{\bf a} Phase diagrams of the GKSL [$(\Gamma$, $J$, $\Gamma'$, $J_{\rm R})/|K| = (0.1, 0, 0, 0)$], PKSL8 [$(0.4,0,-0.05,0.2)$], PKSL14 [$(0.3,0,0,0)$], and $\pi$-flux QSL [$(1,0,0.3,0)$] in magnetic fields applied in and out of the honeycomb plane \cite{JCWang2019,JCWang2020,JCWang2023}. The notation ``6-cone'' (``16-cone'') refers to a phase whose low-energy spinon dispersion remains nodal with 6 (16) remaining cones. The solid red (open blue) points represent first-order (continuous) phase transitions.
{\bf b} Representation of two Z$_2$ vortices inserted at the hexagons marked by red crosses, which define a branch cut (dot-dashed line).
{\bf c}-{\bf d} Energy spectra of the mean-field Hamiltonian in the vicinity of the gap, computed for a large system in two different external magnetic fields $\pmb B \parallel \hat{n}_c$ and in the presence of two vortices \cite{JCWang2019}. The interaction parameters are those of the PKSL14 and the mean-field parameters are determined by VMC calculations with system size 8$\times$8$\times$2. At $g\mu_B B/|K| = 0.17$ (c), each vortex binds two complex fermion modes plus a Majorana zero mode ($\nu = 5$). At $g\mu_B B/|K| = 0.69$ (d), each vortex binds two mid-gap complex fermion modes ($\nu = 4$).}
\label{PKSL14fields}
\end{figure*}

\subsection{Families of Multinode QSLs}

We have shown for the PSG of the KSL that there exist multiple gapless QSLs with different numbers of cones in the spinon spectrum. In fact the number of cones is restricted by symmetry, which we discuss first for the extended Kitaev model with only nearest-neighbor decoupling terms in the mean-field theory. In this case, cones may only be located on the mirror planes of the BZ, and for any momentum point $\pmb k$ other than K and K$^\prime$ their symmetry group is $\{E, PT, M, C_2 T\}$ (where  $M$ is the mirror), whose coset contains $24/4 = 6$ elements. In general each representation in the coset transforms a cone into a new one, and thus cones located at $\pmb k$ appear in multiples of 6. The points K and K$^\prime$ are invariant under the little co-group $D_{3} \times \{E, PT\}$ and transform into each other under $T$, as a result of which a cone at K can occur only in a pair with a cone at K$^\prime$. The zone center ($\pmb k = 0$) respects the full $D_{3d} \times Z_2^T$ symmetry of the system, which, like the M points, does not allow a cone within the Kitaev PSG. Thus a general gapless QSL with this PSG has a spinon dispersion that contains $6n + 2$ cones, where $n \geq 0$. In our VMC studies we have realized the three cases $n = 0$, 1, and 2 from this family, and also the member $n = 1$ of a different family, the $\pi$-flux QSLs. 

\noindent
{\bf $\pmb{6n + 2}$-cone QSLs.} Within the Kitaev PSG we have found a large family of nodal Z$_2$ QSLs containing the KSL phase (2 Majorana cones), the PKSL8 (8 cones), and the PKSL14 (14 cones). These multinode QSLs are sufficiently stable to become the absolute ground state of extended Kitaev models over a significant region of parameter space. A PKSL20 state was also found in Ref.~\cite{JCWang2019}, although it was not the ground state for any parameters analyzed. The $6n + 2$ symmetry-protected cones in the spinon bands are a striking hallmark of this family, and form the basis for complex gapless spin excitation spectra. 

\noindent
{\bf $\pmb{12m + 6n + 2}$-cone QSLs.} In the presence of further-neighbor interactions, including strong $J_{\rm R}$, the above discussion is generalized: cones may be located at a general momentum $\pmb k$, the symmetry group is $\{E, PT\}$, and its coset contains $24/2 = 12$ elements, meaning that these cones appear in multiples of 12 \cite{Yang24B}. Members of this extended family have yet to be explored.

\noindent
{\bf $\pmb{4(6n + 2)}$-cone QSLs.} Our leading example of a family of nodal Z$_2$ QSLs with a PSG different from the KSL is the family possessing the $\pi$-flux PSG. The uniform $\pi$-flux QSLs have $4(6n + 2)$ cones that form an enhanced periodicity in the spinon dispersion (indicating translation-symmetry fractionalization). In the presence of further interactions, this family is extended to become the set of 4(12m + 6n + 2)-cone QSLs.

\noindent
{\bf Further families.} Because there are 144 different PSGs for Z$_2$ QSLs respecting the symmetry of the Kitaev honeycomb model \cite{YouPSG}, it is reasonable to assert that further multinode Z$_2$ QSLs should be found within the phase space of 2D lattice models with SOC. A fruitful direction of research would be to investigate which interactions can stabilize nodal Z$_2$ QSLs with other PSGs.

\section{Multinode QSLs in applied magnetic fields}
\label{SecV}

One of the most fundamental contributions contained in Kitaev's seminal paper \cite{Kitaev} was its exposition of anyonic excitations arising when the spectrum of the fermions is gapped and characterized by a Chern number $\nu$. In the example of the KSL, a magnetic field applied in any orientation not orthogonal to an Ising axis opens a gap, making the system a CSL with $\nu = \pm 1$. Kitaev proceeded to classify 16 different types of CSL based on the statistics of their Z$_2$ vortex ($\pi$-flux) excitations, which are Abelian anyons when $\nu$ is even and non-Abelian anyons when $\nu$ is odd. Hence the KSL in an applied field is a non-Abelian CSL, and $\nu$ also determines the fusion rules, edge modes, topological spins, and mutual statistics of the anyonic excitations of each CSL. The related experimental quantity is the thermal Hall conductance. Because the Chern number of a CSL has a direct association with the number of field-gapped cones, the existence of multinode QSLs suggests the possibility of realizing a large number of CSL types. 

\noindent
{\bf From multinode QSLs to gapped CSLs with $\pmb B \parallel [1 \, 1 \, 1]$.} In general a magnetic field $\pmb B \parallel [1 \, 1 \, 1] (\parallel \hat{n}_c)$, turns all multinode QSLs into gapped CSLs. To perform VMC calculations in a magnetic field, we add the Zeeman-coupling term, $H_{\rm Zeeman} = g \mu_B {\pmb B} \! \cdot \! \frac12 C^\dag_i {\pmb \sigma} C_i$, to the mean-field Hamiltonian and reoptimize the parameters under Gutzwiller projection. We use VMC with fixed interaction parameters to deduce the phase diagrams of each multinode QSL as a function of the field strength, and show the results in Fig.~\ref{PKSL14fields}a. For the pure FM Kitaev model ($K < 0$), we find that the non-Abelian CSL terminates at a critical field $g\mu_{B}B/|K| = 0.19$, not far from the result $g\mu_{B}B/|K| = 0.144$ of Ref.~\cite{HCJiang}, beyond which it enters a topologically trivial phase connected to the fully polarized state. For the AFM Kitaev model ($K > 0$), the situation remains a subject of intense debate \cite{Zhu2018,Trivedi2019,Jiang2018,Hickey2019,WangQH2020,Zhang2022,TXiang2024,Trivedi2024,ChuanChen2024}, which we do not enter here. In Fig.~\ref{PKSL14fields}a we show the field-induced behavior of a GKSL with weak $\Gamma$, which is very similar to the KSL case.  

In the PKSL14, each of the seven pairs of cones becomes gapped in a $[1 \, 1 \, 1]$ field and contributes a Chern number $\nu = \pm 1$. Figure \ref{PKSL14fields}a shows a very rich field-induced phase diagram in this situation, starting with a $\nu = 5$ CSL at small fields and followed by two weakly first-order phase transitions with increasing $|\pmb B|$, first to a $\nu = 4$ CSL and then directly to a trivial phase. The PKSL14 in a field therefore realizes two of the higher CSLs in Kitaev's 16-fold way, the non-Abelian $\nu = 5$ CSL and the Abelian $\nu = 4$ CSL. 

The topological consequences of these high Chern numbers are well established in theory, but to date have little contact with experiments. To summarize the situation, we show mean-field calculations where we introduce a pair of vortices, $\sigma$ \cite{Kitaev}, as represented in Fig.~\ref{PKSL14fields}b, by reversing the signs of the mean-field terms on every bond cut by the red line, and diagonalize a large system. With periodic boundary conditions the system is wrapped on a torus, and the ground-state degeneracy (GSD) provides an independent statement of the number of topologically distinct quasiparticle types. In the KSL ($\nu = 1$), each vortex binds one Majorana zero mode, and this odd number is the origin of the non-Abelian statistics. The fusion rule for vortices is $\sigma \times \sigma = 1 + \varepsilon$, where 1 denotes the vacuum and $\varepsilon$ the fermion, and these three topologically distinct sectors ($1, \varepsilon, \sigma$) match the GSD of 3, as shown in Table \ref{tab:mass}. 

In the $\nu = 5$ CSL, each vortex traps two complex fermion modes and an unpaired Majorana zero mode, as Fig.~\ref{PKSL14fields}c shows explicitly, but the topological structure of this non-Abelian CSL is the same as the KSL, again with a GSD of 3 (Table \ref{tab:mass}). In the $\nu = 4$ CSL, each vortex traps four Majorana zero modes, which couple to form two complex fermion modes, resulting in a total of four mid-gap modes, none at zero energy (Fig.~\ref{PKSL14fields}d). Here the vortex, $m$ \cite{Kitaev}, is a semion and two vortices form a boson ($m \times m = 1$). The composite of a fermion and a vortex creates a new topological entity, $e = m \times \varepsilon$, which is also a semion and obeys $e \times e = 1$. The GSD in this case is 4, as shown in Table \ref{tab:mass}. At fields beyond the $\nu = 4$ CSL lies the trivial phase with no mid-gap modes. To conclude the description of Table \ref{tab:mass}, any CSL with Chern number $\nu$ has $\nu$ branches of chiral Majorana edge states, giving a total chiral central charge $c_- = \nu/2$. The associated quantized physical observable is the thermal Hall conductance, $\kappa_{xy} = \Lambda T c_-$, where $\Lambda = \pi k_B^2 / 6h$. Intensive experimental efforts to measure this quantity in the candidate Kitaev material $\alpha$-RuCl$_3$ have to date yielded a number of fascinating but contradictory results \cite{Matsuda2018,Matsuda2021,Ong2021,Ong2022,Bruin2022,Zhou2023}, leaving the situation unresolved due to issues presumably related to sample quality, and in particular to stacking disorder \cite{Tanaka2020,Taillefer2022,Kasahara2022,BruinAPL2022,Taillefer2023,Yan2023,Yan2024}. 

\begin{table}[b]
\centering
\begin{tabular}{c c c c c}
\hline
\hline
Parent state    &\quad CSL &\quad $\;$topological sectors $\;$   &\quad $\;$GSD$\;$&\quad $c_{-}$ \\
\hline
\multirow{2}{*}{PKSL14}
         &\quad $\; \nu = 5 \;$   & $\sigma,\varepsilon,1$   &\quad 3      &\quad 5/2 \\
         &\quad $\nu = 4$    &$e, m, \varepsilon,1$          &\quad 4      &\quad 2 \\
\hline         
\multirow{2}{*}{PKSL8}         
         &\quad $\nu =-4$    &$e, m, \varepsilon,1$          &\quad 4      &\quad $-2$ \\
         &\quad $\nu =-2$    &$a, \bar a, \varepsilon,1$     &\quad 4      &\quad $-1$ \\
\hline         
\multirow{2}{*}{$\pi$-flux QSL}
	 &\quad $\nu =-8$    &$e, m, \varepsilon,1$          &\quad 4      &\quad $-4$ \\
	 &\quad $\nu = 4$    &$e, m, \varepsilon,1$          &\quad 4      &\quad 2 \\
\hline	 
KSL             &\quad $\nu = 1$    & $\sigma,\varepsilon,1$        &\quad 3      &\quad 1/2 \\ 
\hline
U(1) Dirac SL   &\quad KL           & $e,1$                         &\quad 2      &\quad 1 \\ 
\hline
\hline
\end{tabular}
\caption{Field-induced CSL states in extended Kitaev models. $\nu$ is the Chern number. KL denotes the Kalmayer-Laughlin state \cite{rkl}. 1 denotes the vacuum, $\varepsilon$ the fermion, and $\sigma$ the vortices in the non-Abelian CSLs ($\nu = 1$ and 5). $e$ and $m$ are the two different types of vortex in the $\nu = \pm 4$ and $\pm 8$ Abelian CSLs, and are both semions. $a$ and $\bar a$ are antiparticles of each other in the $\nu = - 2$ Abelian CSL. GSD is ground-state degeneracy on a torus and $c_-$ is chiral central charge.}
\label{tab:mass}
\end{table}

Returning to Fig.~\ref{PKSL14fields}a, the PKSL8 has four pairs of cones that all become gapped in a $[1 \, 1 \, 1]$ field, and each contributes a Chern number $\nu = \pm 1$. From VMC calculations, it becomes a $\nu = - 4$ Abelian CSL in a small field and a $\nu = - 2$ Abelian CSL at higher fields, before also passing to the trivial phase. The nontrivial topological excitations of the $\nu = - 4$ CSL, $\varepsilon$, $m$, and $e$, are the same as those of the $\nu = 4$ CSL (Table \ref{tab:mass}), but those of the $\nu = - 2$ CSL are classified differently \cite{Kitaev}. 

\begin{figure}[t]
\centering
\includegraphics[width=\linewidth]{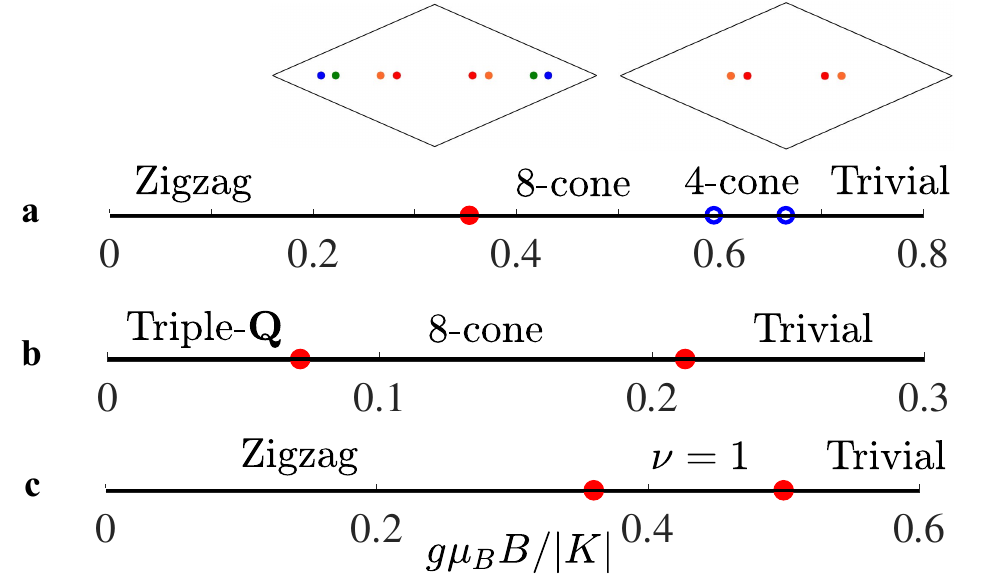}
\caption{Phase diagrams obtained on applying a field to magnetically ordered states. 
{\bf a} In the zigzag phase $(\Gamma$, $J$, $\Gamma'$, $J_{\rm R})/|K| = (1.4,0,-0.05,0)$, a field with $\pmb B \parallel [1 \, \overline{1} \, 0]$ induces a sequence of two gapless states (with 8 cones and 4 cones) after the order is suppressed.
{\bf b} In the triple-{\bf Q} phase $(\Gamma$, $J$, $\Gamma'$, $J_{\rm R})/|K| = (0.4,0,-0.05,0.1)$, a field with $\pmb B \parallel [1 \, \overline{1} \, 0]$ induces one gapless state with 8 cones beyond the ordered phase. 
{\bf c} In the zigzag phase $(\Gamma$, $J$, $\Gamma'$, $J_{\rm R})/|K| = (1.4,0,0,0)$, a field with $\pmb B \parallel [1 \, 1 \, \frac{1}{5}]$ induces one gapped CSL with $\nu = 1$ after the order is suppressed.
Solid red (open blue) points represent first-order (continuous) phase transitions.}
\label{ZZQinField}
\end{figure}

Despite its very large number of cones, the $\pi$-flux QSL does not show a significantly richer field-induced phase diagram. A small $[1 \, 1 \, 1]$ field drives an Abelian CSL phase with $\nu = -8$, the Abelian nature here being a necessary consequence of the fact that the magnetic BZ contains two copies of the compact BZ. This CSL undergoes a second-order phase transition into a $\nu = 4$ CSL at the critical field of ${g\mu_B B/|K|} = 0.14$: despite the quite dramatic change of the Chern number by 12, a continuous transition is protected by the remaining $C_3$ and inversion symmetries in the presence of the field. This system does not enter the trivial phase until a rather high field of ${g\mu_B B/|K|} = 2.3$, which happens with a first-order transition.

\noindent
{\bf Gapless QSLs with $\pmb B \parallel [1 \, \bar1 \, 0]$.} The $[1 \, 1 \, 1]$ field is representative of arbitrary field directions, which give rise to fully gapped CSLs with well defined $\nu$. By contrast, selected high-symmetry field directions preserve the gapless Z$_2$ QSL over a finite range of field magnitudes \cite{c2,JCWang2019,JCWang2020}. We have found this situation for $\pmb B \parallel [1 \, \overline{1} \, 0]$ in all of our primary multinode QSLs (Fig.~\ref{PKSL14fields}a): in the PKSL8 the number of cones is reduced to 4, in the PKSL14 to 6, and in the $\pi$-flux QSL to 16. In each case it is the cones on the high-symmetry line (the horizontal line through K and K$^\prime$) that remain gapless, generally only at very low fields, whereas all other cones become gapped.

\noindent
{\bf Field-induced gapless QSLs from magnetic order.} Mindful of the experimental situation in $\alpha$-RuCl$_3$, we also consider the application of a field to suppress a magnetically ordered phase in favor of a gapless QSL. In Fig.~\ref{ZZQinField}a we show that a high-symmetry, in-plane field, $\pmb B \parallel [1 \, \bar1 \, 0]$, applied to a zigzag ordered ground state (of the $K$-$\Gamma$-$\Gamma'$ model) drives two transitions, the first to a gapless Z$_2$ QSL with 8 Majorana cones and the second, which is continuous, to a 4-cone QSL. This result provides a possible interpretation of the NMR measurement performed on $\alpha$-RuCl$_3$ in Ref.~\cite{YuW2017}, where an in-plane field induced a gapless, disordered state above 7 T and the properties of this state appeared to change above 12 T. Similarly, if one starts in a triple-${\bf Q}$ ordered phase (Fig.~\ref{ZZQinField}b), a $[1 \, \overline{1} \, 0]$ field drives the system into an 8-cone QSL.  

\noindent
{\bf Field-induced CSLs from magnetic order.} In the same spirit, we note that there exists a regime of magnetic field directions for which an ordered magnetic phase is suppressed in favor of a gapped CSL. We illustrate this in Fig.~\ref{ZZQinField}c for the case where a weakly out-of-plane field is applied to the zigzag-ordered phase of Fig.~\ref{Fig_phsdgrm}a.

Quite generally, magnetic fields applied in different directions can have very different physical consequences and we state only that the unexplored parameter space for the discovery of field-induced QSLs remains large. However, our studies do allow us to make one overarching statement. In early work, two of the present authors suggested a field-induced U(1) Dirac QSL state, based on VMC studies of the $K$-$\Gamma$ model without PSG guidance \cite{c2}. By contrast, PSG-guided VMC reveals that U(1) Ans\"atze are always slightly higher in energy than the related Z$_2$ nodal QSL states, and it is generally true that U(1) QSL states are not favored by spin models containing Kitaev-type interactions, or the other interactions produced by strong SOC, presumably because of the discrete and multidirectional nature of their symmetries.

\section{Cone stability and QSL classification}
\label{SecVI}

\noindent
{\bf Stability of Majorana cones.} All nodal Z$_2$ QSLs are obtained by Gutzwiller projection of the fermionic states of a mean-field superconductor; the dispersions of the fermionic spinons contain a discrete number of cones and the low-energy quantum fluctuations of the QSL are described by Z$_2$ gauge fields that couple to the nodal spinons. Usually the form of the low-energy dispersion is not changed by Gutzwiller projection, making the gapped or gapless QSL spectrum agree qualitatively with the spectrum of the mean-field theory: if the cones are robust at the mean-field level, then the number of cones in the corresponding QSL remains unchanged after projection.

Cones in the mean-field theory are protected by the symmetry $PT$. Cones cannot be gapped without breaking $PT$ unless they merge and disappear in pairs, which is, in essence, a second-order quantum phase transition. Thus the number of cones may change only if $PT$ is broken or at a phase transition, and otherwise this protection of the mean-field cones implies that the cones in the spinon dispersion of the QSL are robust against perturbations preserving $PT$.

However, $T$ is broken by a magnetic field applied in an arbitrary direction, or by a three-spin interaction term. Under these circumstances, all the cones become gapped and, if the Z$_2$ gauge field remains deconfined after Gutzwiller projection, the resulting gapped state is a CSL whose properties are determined by the Chern number $\nu$ (Sec.~\ref{SecV}). Every gapped cone contributes $\pm{1\over2}$ to the Chern number, where the sign is given by the sign of the mass of the cone or the chirality of the cone with respect to the perturbation generating the mass. Thus it is clear that the chiralities of all cones must be specified in order to determine the Chern number of the CSL. 

\noindent
{\bf Classification of multinode QSLs.} A complete classification of gapless QSLs is a highly involved and incomplete issue. Here we contribute only by summarizing the above deductions, which constitute a series of necessary conditions. (o) We found that the PSG is necessary, including to describe symmetry fractionalization, but it is not sufficient. (i) We needed to know the number of cones and how they are symmetry-related. (ii) We then found it necessary to determine the chirality under mass-generating perturbations. For completeness we note that (iii) the gauge flux excitation can also carry a fractional quantum number under the symmetry \cite{Hermele2013}, which should be known, and (iv) the anomaly-free condition is required to guarantee that the QSLs being classified can be realized in two-dimensional lattice models \cite{Chen2015,Barkeshli2019}. These and perhaps other conditions are required if a classification scheme is to cover all nodal QSLs, including U(1) Dirac QSLs, different symmetry groups, different 2D lattices, and also higher spins than $S = 1/2$ \cite{Liu2010}. For a more detailed discussion of classification theory, we refer the reader to Ref.~\cite{Liu2020}, where a series of nodal Z$_2$ QSLs was found in the extended $K$-$\Gamma$ model by introducing a spatial anisotropy in the interactions (i.e.~a ``strained honeycomb'' lattice). Extended Kitaev models with four-spin interactions that preserve the local gauge structure and the integrability have been shown to host phases including a uniform $\pi$-flux QSL with a Fermi surface and a $\pi$-flux nodal QSL with 2 cones \cite{Zhang2019,Zhang2020}.

\section{Summary}
\label{SecVII} 

Extended Kitaev models may contain any or all of the Kitaev ($K$), Heisenberg ($J$), off-diagonal ($\Gamma$ and $\Gamma'$), and ring-exchange ($J_{\rm R}$) interactions, which in combination with the field strength ($|B|$) and direction ($\hat{B}$) give a very large parameter space to explore for exotic phases and excitations. We have shown by systematic VMC calculations that the parameter space contains gapless Z$_2$ QSLs with 2, 8, and 14 nodes in the first BZ. While the 2-cone state, the GKSL, has the physics of the KSL, the PKSL8 and PKSL14 states have quite different spinonic quasiparticles and a gapless spin response. Nevertheless, it seems clear that all three states share the same physics, of optimizing the kinetic energy of nearly-free spinons with a minimal number of low-energy modes establishing the long-range correlations (i.e.~a discrete number of cones suffices for this, as opposed to a Fermi surface). The GKSL, PKSL8, and PKSL14 all have the same PSG and belong to a single, large family of multinode Z$_2$ QSLs whose spinon excitation spectra contain $6n + 2$ cones. Separate families of nodal Z$_2$ QSLs exist based on different PSGs, and we discovered a representative of the uniform $\pi$-flux QSLs, whose spinon dispersions contain $4(6n+2)$ cones in the original first BZ.
 
In applied magnetic fields of most orientations, all the cones of these multinode QSLs become gapped to provide a wide choice of CSL states. The states we discuss give rise to CSLs with Chern numbers $\nu = \pm 1$, $\pm 2$, $\pm 4$, $\pm 5$, and $\pm 8$, with the total number of such phases being capped at 8 (up to edge chiral boson modes) \cite{Kitaev}. These CSLs offer explicit realizations of many of the topological concepts and properties introduced by Kitaev in the analysis of his model. For specific high-symmetry field directions, however, the QSLs can remain gapless, with a reduced number of nodes remaining protected by the system symmetries. That a gapless multinode QSL can be induced when the applied field destabilizes a magnetically ordered state offers a route to explain much of the existing experimental data on proximate Kitaev materials.  

We take the view that the role of theory is not only to explain experiments but also to predict where the next generation of experiments can search for exotic physical phenomena. Multinode QSLs lying beyond the GKSL region of extended Kitaev models are truly exotic physics, and thus their existence should be confirmed and their properties explored seriously. From the standpoint of computational physics, while VMC gives access to some of the largest system sizes accessible to numerical methods, it is Ansatz-dependent and therefore not unbiased. However, accurate calculations for multinode Z$_2$ QSLs appear unrealistic by ED and cylinder MPS methods. While intrinsically 2D in nature, tensor-network states are also best suited to describing gapped states obeying area-law entanglement, although here one might establish the existence not of the multinode QSL but of the corresponding field-induced CSL. On the theory side, we reiterate our result that the parameter space of multinode QSL phases is considerably enlarged in the presence of $\Gamma'$ and ring-exchange interactions, which is a promising sign for further theoretical discoveries. For experiment, we take the extremely detailed work on both controlled materials properties and thermal conductivity measurements in $\alpha$-RuCl$_3$ as an indicator that all the ingredients are coming into place for the future realization and characterization of multinode QSL phases in quantum magnetic materials.\\

\medskip
\noindent
{\large {\bf Methods}}

\noindent
{\bf Ring exchange.} For completeness, we mention the treatment of the ring-exchange interaction introduced in Sec.~\ref{SecII} in the spinon representation of Sec.~\ref{SecIII}, where it is expressed simply as $\hat P_{\hexagon} = - \hat\chi_{ij} \hat\chi_{jk} \hat\chi_{kl} \hat\chi_{lm} \hat\chi_{mn} \hat\chi_{ni} - {\rm cyclic}(ijklmn) + {\rm H.c.}$, with indices $i,j,k,l,m,n \in \hexagon$ and $\hat\chi_{ij} = c_{i\uparrow}^\dagger c_{j\uparrow} + c_{i\downarrow}^\dagger c_{j\downarrow}$ being the spinon kinetic term. 

\noindent
{\bf PSG and PSG-guidance.} The clearest example of the PSG is that of the KSL, which is known exactly \cite{YouPSG} and can be read from the Majorana representation: because the $c$-fermion does not hybridize with any of the $b^\gamma$-fermions, any PSG operation leaves it invariant. If $g$ is a symmetry operation and $U(g)$ is the representation of $g$ carried by spin-1/2, the gauge operation, $W_i(g)$, following $g$ should then be $W_i(g) = \pm U(g)$. By applying this framework to the most general spin-orbit-coupled QSL of the $K$-$J$-$\Gamma$ model, one obtains the form \cite{c2,JCWang2019}
\begin{eqnarray}
H_{\rm mf}^{\rm SL} & = &\! \sum_{\langle i,j \rangle \in\alpha\beta(\gamma)} {\rm Tr} \, \Big[ U_{ji}^{(0)} \psi_i ^\dag \psi_j \Big]
 + {\rm Tr} \, \Big[ U_{ji}^{(1)} \psi_i^\dag (iR_{\alpha\beta}^\gamma) \psi_j \Big] \nonumber \\ 
& & + \, {\rm Tr} \, \Big[ U_{ji}^{(2)} \psi_i^\dag \sigma^\gamma \psi_j \Big]
 + {\rm Tr} \, \Big[U_{ji}^{(3)} \psi_i^\dag \sigma^\gamma R_{\alpha\beta}^\gamma \psi_j \Big],
\label{emfh}
\end{eqnarray}
where $R_{\alpha\beta}^\gamma = - \frac{i}{\sqrt{2}} (\sigma^\alpha + \sigma^\beta)$ is a rotation matrix and the quantities $U_{ji}^{(m)}$, with $\gamma$ specified by $\langle i,j \rangle$, are mean-field parameters. By considering the PSG of the KSL, the coefficients $U_{ji}^{(m)}$ in Eq.~(\ref{emfh}) are constrained to the forms 
\begin{equation}\label{eu}
\begin{aligned}
& U_{ji}^{(0)}  =  i \eta_0 + i (\rho_a + \rho_c),  \\ 
& U_{ji}^{(1)}  =  i (\rho_a - \rho_c  + \rho_d) (\tau^\alpha + \tau^\beta) + i \eta_3 (\tau^x + \tau^y + \tau^z), \\ 
& U_{ji}^{(2)}  =  i (\rho_a + \rho_c) \tau^\gamma + i \eta_5 (\tau^x + \tau^y + \tau^z),  \\ 
& U_{ji}^{(3)}  =  i (\rho_c - \rho_a - \rho_d) (\tau^\alpha - \tau^\beta), 
\end{aligned}
\end{equation}
where $\eta_{0,3,5}$ and $\rho_{a,c,d}$ are real \cite{JCWang2019}. The reduction from 16 to 6 mean-field parameters in this case is the practical consequence of the PSG-guided approach. 

\noindent
{\bf VMC calculations.} The fact that the projected states, $|\Psi (\pmb R) \rangle = P_{\rm G} |\Psi_{\rm mf}(\pmb R) \rangle$, and hence the final variational states obtained from VMC, depend so crucially on the choice of the mean-field Hamiltonian, $H_{\rm mf}(\pmb R)$, forced us to perform a systematic selection and comparison of the possible mean-field decoupling channels. By exploiting all symmetries, VMC calculations are possible on lattice sizes up to approximately 400 sites; a systematic investigation of size effects in the extended Kitaev model was performed in Ref.~\cite{JCWang2019}. 

In general, more accurate approximations to the true ground-state energy can be found by adding extra variational parameters to the trial wave function, particularly for magnetically ordered states. A common choice in this regard is a set of two-body Jastrow factors, whose physical eﬀect is to build in additional spin correlations. However, it has been shown that including Jastrow factors in VMC calculations for spin models with strong spin-orbit coupling leads only to very minor changes in predicted phase boundaries \cite{Iaconis2018}, and we obtained similar results in the form of very small ($\mathcal{O}(10^{-4})$) improvements in the ground-state energy. Thus we did not implement a Jastrow-factor approach in our VMC exploration of the model parameter spaces. We stress again that the optimal parameters $\pmb R$ can be used more broadly once they have been determined. An important example is to deduce the spinon dispersion of a QSL state by diagonalizing the corresponding mean-field Hamiltonian on a much larger lattice, and from this dispersion locate very precisely the positions of the nodes in a nodal Z$_2$ (or U(1) Dirac) QSL. 

\medskip
\noindent {\bf Acknowledgements}\\
\noindent We thank Wei Li, Shou-Shu Gong, Xiaoqun Wang, Chenjie Wang, Weiqiang Yu, Jinsheng Wen, Jie Ma, and Yuan Li for discussions and for collaboration on related works. This work was supported by the National Key Research and Development Program of China (under Grant No.~2023YFA1406500 and No.~2022YFA1405301), the National Natural Science Foundation of China (under Grants No.~12374166 and No.~12134020), and the Research Grants Council of Hong Kong (under Grant No.~GRF11300819).

\medskip
\noindent {\bf Correspondence}\\
\noindent Correspondence may be addressed to any one of the authors. 

\medskip
\noindent {\bf Author contributions}\\
J.W., B.N., and Z.X.L. wrote the manuscript together.

\medskip
\noindent {\bf Competing interests}\\
\noindent The authors declare no competing interests.

\bibliography{Ref}

\end{document}